\documentstyle[twocolumn,ucsd,epsf]{article}

\def\roughly#1{\raise.3ex
    \hbox{$#1$\kern-.75em\lower1ex\hbox{$\sim$}}}
\def\be{\begin{equation}}
\def\ee{\end{equation}}
\def\bea{\begin{eqnarray}}
\def\eea{\end{eqnarray}}

\newcommand{\square}{\kern1pt\vbox{\hrule height 0.6pt\hbox{\vrule width
0.6pt\hskip 3pt
   \vbox{\vskip 6pt}\hskip 3pt\vrule width 0.6pt}\hrule height 0.6pt}\kern1pt}

\newcommand{\<}{\!\!\!\!\!}

\begin{document}

\renewcommand{\thepage}{\roman{page}}
\renewcommand{\thebean}{\roman{bean}}
\setcounter{page}{-1}
\setcounter{bean}{0}

\begin{center}
November 1992\hfill   UCSD-PTH 92-40

\vskip .6in

\renewcommand{\thefootnote}{\fnsymbol{footnote}}
\setcounter{footnote}{0}

{\large\bf
Strongly Interacting Higgs Sector in the Minimal Standard
Model\,?\,}\footnote{ \noindent This work was supported by the
U.S. Department of Energy under Contract\newline \indent\indent
DE-FG03-91ER40546.}

\renewcommand{\thefootnote}{\alph{footnote}}
\setcounter{footnote}{0}

\vskip .4in

Karl Jansen,\footnote{E-mail: jansen@higgs.ucsd.edu}
Julius Kuti,\footnote{E-mail: kuti@sdphjk.ucsd.edu}
and Chuan Liu\footnote{E-mail: chuan@higgs.ucsd.edu}

\vskip 0.2in

{\em Department of Physics 0319\\
     University of California at San Diego \\
        9500 Gilman Drive\\
        La Jolla, CA 92093-0319 USA}
\end{center}

\newpage

\mbox{}

\vskip 0.7in

\begin{center}
{\bf Abstract}
\end{center}

\vskip 0.2in

\begin{small}
\begin{quotation}
The triviality Higgs mass bound is studied {\it without} lattice regulator
in the spontaneously broken phase of the four dimensional O(4) symmetric scalar
field theory with quartic self-interaction.
A higher derivative term is introduced in the
kinetic energy of the Lagrangian to keep quantum fluctuations finite while
preserving all the symmetries of the model.
When viewed as a {\it finite} field theory in continuum space-time, some
excitations
of the Higgs field have to be quantized with indefinite metric in the
Hilbert space. It is shown that
the associated ghost particles, which exhibit unusual resonance properties,
correspond
to a complex conjugate pair of Pauli-Villars regulator masses in the euclidean
path integral formulation.
The phase diagram
of the O(4) model is determined in a Monte Carlo simulation which
interpolates between the hypercubic lattice regulator and the higher
derivative regulator in continuum space-time. The same method can be used
to calculate the triviality
Higgs mass bound without lattice artifacts.
In a large-N analysis, when compared with
a hypercubic lattice, we find a relative increase in the triviality bound
of the higher derivative regulator
suggesting a strongly interacting Higgs
sector in the TeV region without detectable dependence on the ghost parameters.
\end{quotation}
\end{small}

\newpage



\renewcommand{\thepage}{\arabic{page}}
\renewcommand{\thebean}{\arabic{bean}}
\setcounter{page}{0}
\setcounter{bean}{0}

\renewcommand{\thefootnote}{\arabic{footnote}}
\setcounter{footnote}{0}


\section{Introduction}

Based on higher derivative regulators in quantum field theory,
we have developed a systematic method to investigate
in continuum space-time the possibility of a strongly interacting Higgs sector
in the minimal Standard Model with a very heavy Higgs particle in the TeV
region. This question, which is directly relevant for the Higgs physics program
of the Superconducting Supercollider Laboratory, has never  been answered in
any reliable fashion, although speculations and phenomenological approximations
are abundant. We report here our first results on the problem.

In the minimal Standard Model with $SU(2)_L \times U(1)$ gauge
symmetry the Higgs sector is described by a complex
scalar doublet $\Phi$ with quartic self-interaction.
The Higgs potential has the well-known form
$
V(\Phi^{\dagger}\Phi) = - {1\over2} m_0^2 \Phi^{\dagger}\Phi +
\lambda_0 (\Phi^{\dagger}\Phi)^2 ~
$
where $m_0^2$ is a mass parameter ($m_H=\sqrt 2 m_0$ in tree approximation)
and $\lambda_0$ designates the quartic coupling
constant.
The four real components of the complex scalar doublet $\Phi$ transform
as a real vector $\phi_\alpha, ~\alpha=1,2,3,4$,
under O(4) symmetry transformations while  the Higgs potential remains
invariant.
In the limit where the gauge couplings and Yukawa couplings are neglected
the Higgs sector is described by the O(4) symmetric action of
the scalar field $\phi_\alpha$ with quartic self-interaction.

In earlier lattice investigations
the triviality upper bound for the Higgs mass
was found at 640 GeV
under some well defined set of conditions \cite{KLS,LW,HJJLNY}  with a
lattice
momentum cut-off at 4 TeV. One can show from the equivalence theorem \cite{LQT}
that lattice artifacts with the 4 TeV cut-off remain hidden to a few percent
accuracy below 1.3 TeV  center of mass energy (the physics reach of the SSC) in
the experimentally relevant $W_LW_L$ cross sections. There has been great
concern that this finding was an artifact of the lattice regulator which breaks
Euclidean invariance. In fact, several attempts have been made to increase the
Higgs mass bound by different lattice structures \cite{BBHN} and improved
lattice actions \cite{GKNZ,HNV}.  Although some increase in the Higgs mass
bound was suggested, the underlying  lattice structure remained the major
limitation.

In this work we replace the lattice regulator with a higher
derivative kinetic term in the Higgs Lagrangian which, according to
conventional thinking, acts as a Pauli-Villars regulator preserving all the
relevant symmetries of the theory. We will investigate the regulator dependence
of the theory without lattice artifacts.
The phase diagram
of the O(4) model will be determined in a Monte Carlo simulation which
interpolates between the hypercubic lattice regulator and the higher
derivative regulator in continuum space-time. The same method can be used
to calculate the triviality
Higgs mass bound without lattice artifacts.
In a large-N analysis\footnote{Variants of the large-N analysis in the study
of the O(N) model were developed earlier\cite{HNV,EW,KLS2}.}, when compared
with
a hypercubic lattice, we find a relative increase in the triviality bound
of the higher derivative regulator
suggesting a strongly interacting Higgs
sector in the continuum without detectable dependence on the ghost parameters.

In our first attempt to drive the Higgs mass bound into the strong
interaction region without lattice regulator
we study the O(4) symmetric
scalar field theory in four dimensions with higher derivative regulator.
Consider the euclidean action of the O(4) model,
\begin{equation}
S_E  = \int d^4x \Bigl[ - {1\over2} ~
\vec\phi~(\square + {1\over M^4}\square^3)~\vec\phi
- {1\over2} m_0^2~\vec\phi\cdot\vec\phi + {\lambda_0}
(\vec\phi\cdot\vec\phi)^2 ~ \Bigr] ~ ,
\label{eq:SEO4}
\end{equation}
where the scalar field $\vec\phi$ has 4 components and $\square$ is the
euclidean Laplace operator in four dimensions. The bare parameters
$m_0^2$, $\lambda_0$ are the same as in the ordinary O(4) theory.
The higher
derivative term $M^{-4}\square^3$ acts as a Pauli-Villars regulator
with mass parameter $M$. It represents a minimal modification of the
continuum model,
if we want to render the field theory and its euclidean path integral,
\begin{equation}
         Z = \int [d\vec\phi]~{\rm exp} ~ \Bigl\{ -S_E[\vec\phi] \Bigr\}\; ,
\label{eq:ZE1}
\end{equation}
finite.

To understand the physics of higher derivative regulator effects
we have to derive the euclidean path integral from the Hamiltonian formulation
in the Hilbert space of quantum states. We also have to understand
the spectrum of the Hamiltonian, the role of expected ghost
states in scattering processes, and the  closely related issues of
unitarity and microscopic violations of causality. A simple quantum mechanical
model
illustrates the strategy of our calculations.

\section{Higher Derivative Quantum Mechanics}

Consider the classical
Lagrangian of a non-relativistic particle moving in one dimension:
\begin{equation}
L= \frac{1}{2}(1\!+\!\frac{2\omega^2}{M^2}
{\rm cos}2\Theta)\dot x^2\!
    -\!\frac{1}{M^2}({\rm cos}2\Theta\!+\!\frac{\omega^2}{2M^2})\ddot x^2
    + \frac{1}{2M^4}{\stackrel{\textstyle ...}{x}}^2 - \frac{1}{2}\omega^2x^2
\;.
\label{eq:QMLag}
\end{equation}
Eq.~\ref{eq:QMLag} describes a simple harmonic oscillator with second
and third derivative terms added to the original Lagrangian;
$M$ and $\omega$ are measured
in units of the Newton mass $m$ of the oscillator particle which is set to
$m=1$ for convenience.
The
coefficients of the derivative terms are given in terms of $M$ and an angle
parameter $\Theta$ for simple
interpretation of the results.
The only restrictions imposed are $\omega^2/M^2
< 1$ and $0 \! < \! \Theta \! < \! \pi/2$.

The variational principle
leads to the higher order Euler-Lagrange equation of motion,
\begin{equation}
\frac{1}{M^4}\frac{d^6 x}{dt^6} +
\frac{2}{M^2}\Bigl({\rm
cos}2\Theta\!+\!\frac{\omega^2}{2M^2}\Bigr)\frac{d^4x}{dt^4} ~-
  \Bigl(1\!+\!\frac{2\omega^2}{M^2}
{\rm cos}2\Theta\Bigr)\frac{d^2x}{dt^2} + \omega^2 x = 0 ~,
\label{eq:EulLag}
\end{equation}
which differs from Newton's equation by higher derivative terms with small
coefficients in the limit $\omega^2/M^2 \ll 1$, and $1/M^2 \ll 1$
(limit of large Pauli-Villars regulator masses in field theory).
Even very small higher derivative
terms will have a qualitative impact on the theory as shown in the
following analysis.

The canonical formalism for higher derivative theories was developed first
by Ostrogradski \cite{O}. Three independent generalized coordinates
$x$, $\dot x$, and $\ddot x$  are introduced and the
generalized canonical momenta are defined by
\begin{eqnarray}
\< \pi_x \< &=& \< \frac{\partial L}{\partial\dot x} -
\frac{d}{dt}  \frac{\partial L}{\partial \ddot x} +
\frac{d^2}{dt^2}\frac{\partial L} {\partial {\stackrel{\textstyle ...}{x}}}
= \nonumber \\
& & \!\! \Bigl(1\!+\!\frac{2\omega^2}{M^2}{\rm cos}2\Theta\Bigr)\dot x
+
\frac{2}{M^2}\Bigl({\rm
cos}2\Theta\!+\!\frac{\omega^2}{2M^2}\Bigr){\stackrel{\textstyle ...}{x}}
+ \frac{1}{M^4}\frac{d^5x}{dt^5}\; , \nonumber \\
\nonumber \\
\< \pi_{\dot x}\< &=&\< \frac{\partial L}{\partial\ddot x}\! -\!
\frac{d}{dt}  \frac{\partial L}{\partial {\stackrel{\textstyle ...}{x}}} =
-  \Bigl(\frac{2{\rm cos}2\Theta}{M^2} + \frac{\omega^2}{M^4}\Bigr)\ddot x
- \frac{1}{M^4}\frac{d^4x}{dt^4}, \nonumber \\ \nonumber \\
\pi_{\ddot x}\< &=& \frac{\partial L}{\partial {\stackrel{\textstyle
...}{x}}} = \frac{1}{M^4}{\stackrel{\textstyle ...}{x}} \; .
\label{eq:can_mom}
\end{eqnarray}
With Ostrogradski's method we find the Hamiltonian
\begin{eqnarray}
\lefteqn{
\<\<\< H  = \pi_x\cdot\dot x + \pi_{\dot x}\cdot\ddot x +
\frac{1}{2}M^4\pi_{\ddot x}^2 + \frac{1}{2}\omega^2 x^2
}
\nonumber \\
& &\mbox{}
-\frac{1}{2}\Bigl(1+2\frac{\omega^2}{M^2}{\rm cos}2\Theta\Bigr)\dot x^2 +
\frac{1}{2}\Bigl(\frac{\omega^2}{M^4}+\frac{2{\rm cos}2\Theta}{M^2}\Bigr)\ddot
x^2  . \label{eq:hamiltonian}
\end{eqnarray}
In quantum mechanics the operators of the canonical variables satisfy
standard Heisenberg commutation relations,
\begin{equation}
[ \hat{x}, \hat{\pi}_x ] = i\hbar ~, ~~~~~
[ \hat{\dot x} , \hat{\pi}_{\dot x} ] = i\hbar ~, ~~~~~
[ \hat{\ddot x} , \hat{\pi}_{\ddot x} ] = i\hbar \; .
\label{eq:COMMUTATOR}
\end{equation}
All the other commutators of the six operators vanish. We define the
scalar product in the Hilbert
space with indefinite metric \cite{P},
\begin{equation}
\langle x', \dot x', \ddot x' | x, \dot x, \ddot x \rangle =
\delta (x'-x)~\delta (\dot x' + \dot x)~\delta (\ddot x'-\ddot x) ~,
\label{eq:metric}
\end{equation}
as indicated by the {\it plus sign} in the second $\delta$-function on the
right side of Eq.~\ref{eq:metric}.
The wavefunctions $\psi(x,\dot x, \ddot x) =$
$\langle x, -\dot x, \ddot x |\psi\rangle$
depend on the three independent generalized coordinates.
The completeness relation in the indefinite metric Hilbert space is given by
\begin{equation}
\int\!\!\! dx\!\!\! \int \!\!\! d\dot x\!\!\! \int\!\!\! d\ddot x
\; |x, -\dot x, \ddot x\rangle\langle x, \dot x, \ddot x | = {\rm I} \; .
\end{equation}
The representation of the self-adjoint operators $\hat{x}, \hat{\pi}_x$,
$\hat{\ddot x},  \hat{\pi}_{\ddot x}$
follows the rules of ordinary quantum mechanics,
$\hat{x}~\psi = x\cdot\psi$, $\hat{\pi}_x~\psi =
-i~\frac{\partial\psi}{\partial x}$, and similarly,
$\hat{\ddot x}~\psi = \ddot x\cdot\psi$, $\hat{\pi}_{\ddot x}~\psi =
-i~\frac{\partial\psi}{\partial \ddot x}$.
However, $\hat{\dot x}$ and $\hat{\pi}_{\dot x}$
(also self-adjoint with respect to the indefinite metric)
are represented by
$\hat{\dot x}~\psi = i\dot x\cdot\psi$, and $\hat{\pi}_{\dot x}~\psi =
- \frac{\partial\psi}{\partial \dot x}$.

The euclidean partition function in the presence of external sources is defined
by
\begin{eqnarray}
Z_T(J_0,J_1,J_2) &=& {\rm Tr} e^{-\frac{1}{\hbar} T\, H}  \nonumber \\
&=& \int\!\!\! dx\!\!\! \int \!\!\! d\dot x\!\!\!
\int\!\!\! d\ddot x \; \langle x, -\dot x, \ddot x|
e^{-\frac{1}{\hbar} T\, H}|x, \dot x, \ddot x \rangle \; ,
\label{eq:partf}
\end{eqnarray}
where the time dependent source term $J_0\hat{x} + J_1\hat{\dot x} + J_2\hat{
\ddot x}$ is included in the Hamiltonian.
The partition function of Eq.~\ref{eq:partf} can be written as a Hamiltonian
path integral in six canonical variables. After integrating over
$\dot x, \ddot x, \pi_x, \pi_{\dot x}, \pi_{\ddot x}$, we find
\begin{eqnarray}
\<\<\<\<\<& &\<\< Z_T(J_0, J_1, J_2) = \int d[x(\tau)]~{\rm exp} \Biggl\{
-\frac{1}
{\hbar} \int^T_0 d\tau \biggl(
L_E + J_0(\tau) x(\tau)  \nonumber \\
& & \mbox{} ~~~~~~~~ + J_1(\tau)\frac{dx}{d\tau} +
J_2(\tau)\frac{d^2x}{d\tau^2}\; \biggr) \Biggr\} ,
\label{eq:ZT}
\end{eqnarray}
where $L_E$ is the {\it euclidean} Lagrangian of the higher derivative model:
\begin{eqnarray}
\lefteqn{
\<\<\<\< L_E = \Bigl( \frac{1}{2} + \frac{\omega^2}{M^2}{\rm cos}
2\Theta\Bigr)\Bigl( \frac{dx}{d\tau}\Bigr)^2 + \frac{1}{2}\omega^2x^2(\tau)
} \nonumber \\
& &\mbox{} +
\frac{1}{M^2}\Bigl( {\rm cos}2\Theta +
\frac{\omega^2}{2M^2}\Bigr)\Bigl(\frac{d^2x}{d\tau^2}\Bigr)^2  +
\frac{1}{2}M^{-4}\Bigl(\frac{d^3x}{d\tau^3}\Bigr)^2 ~.
\label{eq:LE}
\end{eqnarray}
In the T $\rightarrow \infty$ limit, after the evaluation of the path integral
in Eq.~\ref{eq:ZT}, we obtain
\begin{eqnarray}
\lefteqn{
\<\<\<\< Z_{\infty}(\tilde{J}_0(E), \tilde{J}_1(E), \tilde{J}_2(E)) =
Z_{\infty}(0,0,0)
} \nonumber \\
 &\times&
{\rm exp} \Biggl\{ -\frac{1}{2} \int dE\tilde{J}(E)\widetilde{D}(E)
\tilde{J}(-E) \Biggr\} ~ ,
\label{eq:Zinf}
\end{eqnarray}
where $\tilde{J}(E)=\tilde{J}_0(E) - i E\tilde{J}_1(E) - E^2\tilde{J}_2(E)$.
The Fourier transform of the quantum mechanical propagator $D(\tau)$ is
given by
\begin{equation}
\widetilde{D}(E) =
\frac{M^4}{(E^2+\omega^2)(E^2+M^2e^{-2i\Theta})(E^2+M^2e^{2i\Theta})}~~.
\label{eq:D(E)}
\end{equation}

The physical significance of the multiple pole structure in the
propagator becomes clear
from the spectrum of the Hamiltonian which can be
transformed to a diagonal form,
\begin{equation}
H =
\omega a^{(+)}a^{(-)} + Me^{-i\Theta}b^{(+)}b^{(-)}
+ Me^{i\Theta}c^{(+)}c^{(-)}
 + \! \frac{1}{2} (\omega + M e^{-i\Theta} + M e^{i\Theta}) \; .
\end{equation}
The creation operators $a^{(+)}, b^{(+)}, c^{(+)}$ and the annihilation
operators $a^{(-)}, b^{(-)}, c^{(-)}$ are linear combinations
of the six canonical variables $x, \dot x, \ddot x, \pi_x , \pi_{\dot x},
\pi_{\ddot x}$ with commutation relations
$[ a^{(-)}, a^{(+)} ] =1$, $[ b^{(-)}, b^{(+)} ] =1$,
$[ c^{(-)}, c^{(+)} ] =1$.
All the other commutators vanish. The adjoint
of $a^{(+)}$ is $\overline{a^{(+)}} = a^{(-)}$, and similarly,
$\overline{b^{(+)}} =
c^{(-)}$, $\overline{c^{(+)}} = b^{(-)}$.

The eigenstates of the Hamiltonian are given by
\begin{equation}
|l,m,n \rangle =
\frac{1}{\sqrt{n!}}(c^{(+)})^n\frac{1}{\sqrt{m!}}(b^{(+)})^m
\frac{1}{\sqrt{l!}}(a^{(+)})^l\; |0,0,0\rangle \;
\end{equation}
where the ground state $|0,0,0\rangle$ is annihilated by $a^{(-)}$,
$b^{(-)}$, $c^{(-)}$. The complex energy eigenvalues
\begin{eqnarray}
E_{l,m,n} \!=\! ( l + \frac{1}{2} )\omega + ( m + \frac{1}{2} ) Me^{-i\Theta}
\!+\! ( n + \frac{1}{2} ) Me^{i\Theta}
\end{eqnarray}
satisfy the symmetry relation $E_{l,m,n} = E^{*}_{l,n,m}$ . In addition
to the ordinary oscillator
excitations we have two conjugate ghost oscillators with complex energies,
in agreement with the two new degrees of freedom in the oscillator.
The energy eigenstates have indefinite metric normalization,
\begin{equation}
\langle l',m',n' |l,m,n \rangle = \delta_{l'l}\delta_{m'n}\delta_{n'm} \; .
\label{eq:norm}
\end{equation}
For $m=n$, the ghost oscillator and its complex conjugate partner are excited
in pairs. The corresponding state vectors in the indefinite metric Hilbert
space have positive norm with real energy. However, unbalanced ghost
excitations
with $m\neq n$ correspond to zero norm state vectors with complex energies.
Some linear combinations of ghost excitations with $m\neq n$ will have negative
norm.
%
%

By repeated differentiation of $Z_T(J_0,J_1,J_2)$ with respect to the three
source terms we can derive the euclidean correlation functions of the variables
$x(\tau)$, $\dot x(\tau)$, $\ddot x(\tau)$. To compare analytic
calculations
with computer simulation results (Fig.~1)
we discretize the euclidean path integral with 100
time slices, $T= 100~\Delta\tau$. The time slice is scaled out  by the
choice $\Delta\tau = 1$ for convenience, so that
dimensional quantities are measured in $\Delta\tau$ units. The
parameters  $\omega =0.1$, $M=0.5$, $\Theta=1.4~$ (radian) were selected for
illustration.  Fig.~1a depicts $D_\omega (\tau) = \langle \ddot x(\tau)
\ddot x(0) +
2 M^2{\rm cos}2\Theta\dot x(\tau)\dot x(0) +
M^4 x(\tau)x(0) \rangle$ which isolates the $\omega$-pole
in the Fourier transform $\widetilde{D}(E)$ of the propagator
$D(\tau) = \langle x(\tau) x(0) \rangle$
by eliminating the ghost pair.
In other words, $D_\omega (\tau)$
is the propagator of the ordinary
oscillator without higher derivatives. Fig.~1b shows the correlation function
$D_M (\tau) = \langle \dot x(\tau)\dot x(0) +
\omega^2  x(\tau) x(0) \rangle$ which isolates the complex
conjugate ghost pair in the propagator by eliminating the $\omega$
-pole. It oscillates, and the exponential decay of its amplitude is
controlled by M. The field theory analogues of these correlation functions
plays
an important role in our analysis of Higgs mass calculations and ghost effects
in
computer simulations.

\section{Higher Derivative Field Theory}

The quantization procedure for the higher derivative Lagrangian,
\begin{equation}
L = \frac{1}{2} \Bigl( \partial_\mu \vec\phi\partial^\mu\vec\phi
- m^2_0\vec\phi\cdot\vec\phi \Bigr)
 + \frac{1}{2M^4}~\square\partial_\mu\vec\phi~\square\partial^\mu\vec\phi
- \lambda_0\Bigl( \vec\phi\cdot\vec\phi\Bigr) ^2 ~,
\label{eq:O(4)_Lagr}
\end{equation}
in Minkowski space-time is very similar to the quantum mechanical example.
There are three independent sets of field variables (generalized coordinates)
$\phi_\alpha (\vec x, t)$, $\dot{\phi}_\alpha (\vec x, t)$, and
$\ddot{\phi}_\alpha (\vec x, t)$, $\alpha=1,2,3,4$. The canonical
field momentum variables $\Pi_{\phi_\alpha}$, $\Pi_{\dot{\phi}_\alpha}$,
$\Pi_{\ddot{\phi}_\alpha}$ are defined in close analogy with
Eq.~\ref{eq:can_mom}.
The Heisenberg commutators of the canonical
variables have the same form as in ordinary quantum field theory with
positive definite metric, in accord with Eq.~\ref{eq:COMMUTATOR}.
The state vectors $|\psi\rangle$ are described by wave functionals
in the field-diagonal representation of the indefinite metric Hilbert space,
\begin{equation}
\psi\Bigl\{ \phi_\alpha (\vec x), \dot{\phi}_\alpha (\vec x),
\ddot{\phi}_\alpha (\vec x) \Bigr\} =
\langle \phi_\alpha (\vec x), - \dot{\phi}_\alpha (\vec x),
\ddot{\phi}_\alpha (\vec x) | \psi \rangle ~.
\end{equation}
The operators $\hat{\phi}_\alpha$, $\hat{\Pi}_{\phi_\alpha}$,
...  are represented according to our strategy of higher derivative
quantum mechanics:
\begin{equation}
\hat{\phi}_\alpha (\vec x)~\psi\Bigl\{ \phi_\alpha (\vec x),
\dot{\phi}_\alpha (\vec x),
\ddot{\phi}_\alpha (\vec x) \Bigr\} =
{}~~~~~\phi_\alpha (\vec x)\cdot\psi\Bigl\{ \phi_\alpha (\vec x),
\dot{\phi}_\alpha (\vec x),
\ddot{\phi}_\alpha (\vec x) \Bigr\} ~,
\end{equation}
\begin{equation}
\hat{\Pi}_{\phi_\alpha}~\psi\Bigl\{ \phi_\alpha (\vec x),
\dot{\phi}_\alpha (\vec x),
\ddot{\phi}_\alpha (\vec x) \Bigr\} =
- i \frac{\delta}{\delta \phi_\alpha (\vec x)} ~
\psi\Bigl\{ \phi_\alpha (\vec x), \dot{\phi}_\alpha (\vec x),
\ddot{\phi}_\alpha (\vec x) \Bigr\} ~,
\end{equation}
with similar representation for the
$\hat{\ddot{\phi}}_\alpha$, $\hat{\Pi}_{\ddot{\phi}_\alpha}$ pair.
However, the representation of the
$\hat{\dot{\phi}}_\alpha$, $\hat{\Pi}_{\dot{\phi}_\alpha}$ pair reveals the
indefinite metric:
\begin{equation}
\hat{\dot{\phi}}_\alpha (\vec x)~\psi\Bigl\{ \phi_\alpha
(\vec x), \dot{\phi}_\alpha (\vec x),
\ddot{\phi}_\alpha (\vec x) \Bigr\} =
i~\dot{\phi}_\alpha (\vec x)\cdot\psi\Bigl\{ \phi_\alpha (\vec x),
\dot{\phi}_\alpha (\vec x),
\ddot{\phi}_\alpha (\vec x) \Bigr\} ~,
\end{equation}
\begin{equation}
\hat{\Pi}_{\dot{\phi}_\alpha}~\psi\Bigl\{ \phi_\alpha (\vec
x), \dot{\phi}_\alpha (\vec x),
\ddot{\phi}_\alpha (\vec x) \Bigr\} =
-  \frac{\delta}{\delta \dot{\phi}_\alpha (\vec x)} ~
\psi\Bigl\{ \phi_\alpha (\vec x), \dot{\phi}_\alpha (\vec x),
\ddot{\phi}_\alpha (\vec x) \Bigr\} ~.
\end{equation}
Following the procedure we developed for higher derivative quantum mechanics
it is straightforward to derive the anticipated euclidean path integral
of Eq.~\ref{eq:ZE1} with the euclidean action $S_E$ given by
Eq.~\ref{eq:SEO4}. Details will be given elsewhere.

The system defined by the partition function has
two phases, as expected.
In the symmetric phase of the model
we find the original massive particle (with four components in the
intrinsic O(4) space), and we also find a complex ghost pair
with intrinsic
O(4) symmetry whose mass scale is set by the Pauli-Villars
mass parameter $M$.
In the broken phase we find a Higgs
particle with mass $m_H$, and 3 massless Goldstone excitations with residual
O(3) symmetry. In addition,
the 4-component heavy ghost particle and its complex conjugate partner
also appear in the excitation spectrum of the
broken phase.
%
%

In computer simulations (Fig.~2) we have
to discretize the already regulated and finite theory: the
continuum operators $\square$ and $\square ^3$ are replaced in
Eq.~\ref{eq:SEO4} by the equivalent hypercubic lattice operators. The
lattice spacing $a$ defines a new short distance scale in the theory with the
associated lattice momentum cut-off at $\Lambda = \pi/a$.
At fixed $M/m_H$,
the limit $a \to 0$
corresponds to the Pauli-Villars regulated finite theory in the continuum. At
fixed lattice spacing $a$, the $aM \to \infty$ limit will produce the
ordinary O(4) field
theory on a hypercubic lattice.

\section{The Higgs Mass Bound and Large-N}

Perhaps the most distinguished feature of four-dimensional scalar field
theories is the necessity to suppress quantum
fluctuations at short distances in
the quartic interaction. Without this suppression even an infinitely strong
bare coupling constant $\lambda_0$ would be screened completely. The screening
effects produce a non-interacting (trivial) theory with vanishing
renormalized coupling which is experimentally not acceptable within the
framework of the minimal Standard Model.
The Higgs mass (inverse
correlation length) has to be kept finite in cut-off units.
%
%

To illustrate how this scenario works, the Higgs mass bound will be estimated
in the large-N limit as we interpolate between the lattice O(N) model and
the continuum O(N) model with Pauli-Villars regulator.
The mass of the Higgs particle is given by
\begin{equation}
\frac{m_H}{M} = C(aM,\lambda_0)\cdot {\rm exp} \Bigl\{-\frac{16\pi^2
v^2}{N\cdot m_H^2} \Bigr\}
\label{eq:C_PV}
\end{equation}
where the amplitude $C(aM,\lambda_0)$ is calculable from a Goldstone
loop diagram. At fixed $M/m_H$,
$C_{PV}(\lambda_0) = \lim_{\, aM\to 0} C(aM,\lambda_0)$
is the higher derivative (Pauli-Villars) regulator limit of the amplitude
in continuum
space-time. At fixed $a$, $C_{LAT}(\lambda_0) = \lim_{\, aM\to \infty}
[aM \cdot C(aM,\lambda_0)]$
corresponds to
the original hypercubic lattice theory without higher derivative
regulator term (the reason for the prefactor $aM$ in front of the
amplitude $C(aM,\lambda_0)$ is the replacement of $m_H/M$ by $a\cdot m_H$
on the left side of Eq.~\ref{eq:C_PV}  in the $M\to \infty$ limit).
Numerically both amplitudes are the largest in the $\lambda_0 \to
\infty$ limit where we find $C_{PV}= e^{1/4}$ and $C_{LAT}=e^{2.896}$,
for $N=4$ .

We find an increase in the Higgs mass bound  by comparing the
ratio  $m_H/v$ in the two different schemes,
\begin{equation}
{(m_H/v)_{PV}\over (m_H/v)_{LAT}} = \sqrt{ {{\rm ln}\, (am_H)_{LAT} + {\rm
ln}\,
C_{LAT}(\lambda_0) \over {\rm ln}\, (M/m_H)_{PV} + {\rm ln}\,
C_{PV}(\lambda_0) }} ~~.
\label{eq:RATIO}
\end{equation}
In the $aM \to \infty$ limit, $m_H\cdot a = 0.5$ corresponds to
few percent lattice effects (deviation from $R=1$ in Fig.~3b)
in the physical cross section
of elastic $W_LW_L$ scattering at a center of mass energy $W = 2m_H$.
At fixed $M/m_H = 4$,  in the $a \to 0$ Pauli-Villars limit, we find a
comparable few percent deviation from
$R=1$
(ghost effects) in the $W_LW_L$ cross section at $W = 2 m_H$.
Using large-N numbers for $C_{LAT}$ and
$C_{PV}$ at $\lambda_0 = \infty$, we find from Fig.~3a that
the Higgs mass bound increases from $m_H  \approx 3v$ to
$m_H \approx 5v$ as we interpolate from the hypercubic lattice action
to the continuum
higher derivative regulator. If the trend of the relative comparison
remains the same in the O(4) model, the largest
Higgs mass with higher derivative regulator without ghost effects will be
driven into the TeV range.

\section{Conclusions}

We derived the Pauli-Villars regulated euclidean quantum field theory
for scalar fields with quartic self-interaction as the quantum
mechanical theory of higher derivative Lagrangians in indefinite metric
Hilbert space.
The ghost states with zero norm were identified and their effect on the Higgs
mass bound was studied in the large-N approximation to the O(4) model.
The increase in the Higgs mass bound was calculated in
the minimal standard model with its {\em two} independent parameters
$m_H$ and $v$. Scattering amplitudes were restricted to the energy range
below the mass of the conjugate ghost pair where dependence on $M$ is almost
negligible.

We expect
that the S-matrix remains unitary in physical processes at arbitrary
center of mass energies, as it was suggested
in quantum electrodynamics earlier \cite{LEEWICK}. It will be very interesting
to study the physical structure of scattering amplitudes in the energy range
of the complex ghost pair, and beyond.
With our understanding of the underlying mathematical structure of the
higher derivative
quantum theory, we hope to return
to these questions in the near future with a more detailed presentation.

The non-perturbative computer investigation of the higher derivative Lagrangian
required the introduction of an underlying hypercubic lattice structure.
It is important to emphasize that we work in the large $\Lambda /M$ limit where
lattice effects are negligible compared with ghost effects, so that
it is the regulated continuum theory which is simulated.
We reported here the first results on the phase diagram of the O(4) model and
results on the Higgs mass bound are expected soon.

\subsection*{Acknowledgements}

We thank Peter Hasenfratz and Aneesh Manohar for useful conversations and
comments.
This work was supported by the DOE under contract DE-FG03-91ER40546.

\newpage

\section*{Figure Captions}

\begin{description}

\item[Fig.~1:]
Solid lines represent the analytic form
of the correlation functions of the higher derivative oscillator model.
The data were generated in a high precision computer simulation (the
error bars in (b) are smaller than the size of the marks representing
the data points).

\item[Fig.~2:]
The phase diagram of the interpolated Pauli-Villars lattice model
at infinite bare coupling is shown (the lattice hopping parameter
$\kappa$ originates
from the discretization of the euclidean Laplace operator). The dotted
line is the analytic form of the critical line in the large-N approximation.
The data points of critical $\kappa$ values are simulation results in
the O(4) model.
The solid line displays the fixed $M/m_H=10$ ratio towards the $aM \to 0$
Pauli-Villars limit of the higher derivative continuum theory,
with $\kappa$ values scaled by a factor
of five for convenience.

\item[Fig.~3:]
(a) Arrows mark
the large-N Higgs mass bounds with Pauli-Villars and lattice
regulators, respectively,
as $m_H/v$ is plotted against $\xi_L = am_H$, or $\xi_{PV} = M/m_H$
at $\lambda_0 = \infty$; (b) deviations from
$R=1$ measure regulator effects in the $W_LW_L$ cross section.

\end{description}

\pagebreak

\begin{figure}[t]
\centerline{\epsfbox{fig1.ps}}
\end{figure}

\begin{figure}[t]
\centerline{\epsfbox{fig2.ps}}
\end{figure}

\begin{figure}[t]
\centerline{\epsfbox{fig3.ps}}
\end{figure}

\end{document}